\documentclass[10pt]{wlscirep}
\usepackage[utf8]{inputenc}
\usepackage[T1]{fontenc}
\usepackage{amsmath}
\usepackage{graphicx}
\usepackage{tikz}

\title{Trans-stenotic pressure drop estimation from PC-MRI and ultrasound imaging velocimetry using a modified Bernoulli equation}

\author[1,2]{Ali Amiri}
\author[1]{Johan T. Padding}
\author[2,*]{Selene Pirola}
\author[1,*]{Willian Hogendoorn}
\affil[1]{Delft University of Technology, Complex Fluid Processing (Process $\&$ Energy), Delft, 2628 CB, The Netherlands}
\affil[2]{Delft University of Technology, Cardiovascular Biomechanics (BioMechanical Engineering), Delft,  2628 CD, The Netherlands}

\affil[*]{S.Pirola@tudelft.nl (Selene Pirola), W.J.Hogendoorn@tudelft.nl (Willian Hogendoorn)}
\affil[*]{Senior authors contributed equally to this work}

\begin{abstract}

Accurate non-invasive estimation of trans-stenotic pressure drops remains a challenge. In clinical practice, pressure drops are often estimated from velocity measurements using Bernoulli-based formulas, but these simplified relations do not explicitly account for how pressure losses change with the flow regime. Here, we introduce a modified Bernoulli (MB) formulation that incorporates regime-dependent pressure losses through a Reynolds-number-dependent loss coefficient. Steady in-vitro experiments were performed in an idealized stenosis model over physiologically relevant flow rates ($0.65-3.9$ L/min), combining direct pressure measurements with ultrasound imaging velocimetry (UIV) and phase-contrast magnetic resonance imaging (PC-MRI) to measure velocities. The MB model was calibrated from the measured pressure drops and then evaluated against the simplified Bernoulli (SB) and extended Bernoulli (EB) formulations. Over the tested flow regime, MB agreed best with the measurements (typically within about $\pm 10\%$). SB and EB showed larger biases, with errors of roughly $10-55\%$ (SB) and -$15-25$$\%$ (EB), and overestimated the pressure drop in the clinically relevant range. We additionally quantified the effect of PC-MRI in-plane pixel size on MRI-based pressure estimates. Coarse in-plane pixel size of $1.33$ mm, corresponding to only approximately $3-4$ pixels across the $3.3$ mm throat radius of the $6.6$ mm stenosis throat, led to systematic underestimation of flow rate and bulk velocity (about -$34$ to -$44\%$) and, consequently, of the MB-predicted pressure drop (about -$52$ to $-62\%$). In contrast, the peak throat velocity was substantially less sensitive to pixel size, resulting in smaller estimation errors (about -$13$ to -$18.7\%$) when used as input for the MB. Overall, the results demonstrate that accounting for flow-regime-dependent loss mechanisms enhances pressure drop estimation, and that sufficient sampling of the stenotic throat is crucial for MRI-based flow rate and pressure drop estimation. In addition, peak-velocity-based MB pressure drop estimations are less sensitive to pixel size.

\end{abstract}
\begin{document}

\flushbottom
\maketitle
\thispagestyle{empty}

\section*{Introduction}

Over the past few decades, cardiovascular diseases have become one of the top three causes of death around the world, with arterial stenosis ranking among the most significant contributors to global cardiovascular disease \cite{Thiriet:2015Stenosis}. Stenosis is a common disease, affecting up to 5$\%$ of heart valves \cite{Lindman:2016} and 10$\%$ of blood
vessels \cite{Messas:2020}. It leads to pathological pressure drops,  which increase
the heart workload and cause cardiovascular complications and mortality \cite{Ross:1976}. Therefore, pressure drops serve as crucial indicators of stenosis severity.

% Introduction
One of the major challenges in the management of stenosis is the accurate assessment of lesion severity \cite{Paraskevas:2024}. The hemodynamic significance could be assessed by measuring the pressure drop across the narrowed region. However, accurate determination of trans-stenotic pressure drops represents a critical challenge in cardiovascular medicine, directly influencing therapeutic decisions for patients with arterial stenoses and valvular disease. While catheter-based measurements remain the clinical gold standard, their invasive nature and associated risks have motivated extensive research into non-invasive alternatives \cite{Ha:2017,Zardkoohi:2019}. The convergence of advanced medical imaging, computational methods, and refined analytical methods has created new possibilities for pressure drop assessment. However, significant challenges persist in translating these approaches to routine clinical practice.

% Bernoulli methods
Simplified formulations of the Bernoulli equation have long dominated clinical practice for non-invasive pressure drop estimation, such as echocardiographic evaluation of valvular stenosis \cite{Anavekar:2009}. The simplified Bernoulli equation (SB) remains widely used, even though its limitations are well documented
\cite{Gill:2023, Kazemi2022, Ha:2017, Paraskevas2:2024, Sakthi:2005, Oshinski:1996}.  Kazemi et al. \cite{Kazemi2022} evaluated relative pressure estimation in a stenotic phantom using 4D-flow MRI  (time-resolved 3D phase-contrast magnetic resonance imaging). They found that the pressure drop estimation from SB could deviate by 22$\%$ at moderate flow and 40$\%$ at the highest flow rate tested. Garcia et al. \cite{Garcia2000} advanced this model by introducing an extended Bernoulli equation (EB). In this model the ratio between the effective orifice area and the vessel area is incorporated. This effectively accounts for the downstream pressure recovery due to the jet deceleration. However, despite this improvement, the EB does not explicitly capture the change in flow physics from laminar, viscous-dominated losses to turbulent, inertia-dominated losses \cite{White:2011}. 
% Ultrasound-based pressure drop estimation
These limitations have motivated alternative image-based approaches for estimating pressure gradients non-invasively. Ultrasound-based pressure gradient estimation has thus advanced beyond the classical Doppler--Bernoulli approach. Doppler echocardiography remains central in clinical stenosis assessment and pressure drops are usually inferred from simplified Bernoulli relations \cite{Anavekar:2009,Baumgartner:2017,Gill:2023}. Recent vector-flow and synthetic-aperture ultrasound methods instead estimate pressure differences from spatially and temporally resolved velocity fields using the Navier--Stokes equation. These approaches have been demonstrated in experimental phantoms, carotid bifurcation models, and in vivo carotid measurements \cite{Olesen:2018,Nguyen:2019,Haslund:2021,Haslund:2025}. However, they require resolved velocity fields, whereas Bernoulli models remain attractive when the aim is to estimate the overall trans-stenotic pressure drop from reduced velocity information.

% MRI: effect of pixel size
4D-flow MRI has become a promising technique for non-invasive quantification of cardiovascular hemodynamics, providing volumetric velocity fields that are increasingly used to characterize stenotic jets, secondary flow structures, and clinically relevant flow-derived biomarkers \cite{Markl:2012, Soulat:2020, Allen:2019}. Markl et al. outline both the clinical potential and the practical limitations of 4D-flow MRI for cardiovascular assessment \cite{Markl:2012}. Multiple studies emphasize that quantitative interpretation is strongly limited by acquisition constraints, most notably pixel size \cite{Dirix:2022, Ha:2017, Cherry:2022}. Larger pixel sizes can systematically bias measured velocities through partial-volume effects (PVE). This bias is amplified by smoothing of sharp gradients, particularly in narrow stenotic throats and high-shear regions. Because any downstream pressure estimate depends directly on the accuracy of the measured velocity, this spatial sampling limitation propagates into pressure drop estimation. This consequence is explicitly quantified in the present study.

Building on spatially resolved velocity data, 4D-flow MRI has also been applied to non-invasive pressure gradient estimation through a range of approaches. Relative pressure fields have been reconstructed from the pressure-Poisson formulations derived from Navier--Stokes equations \cite{Nolte:2021,Saitta:2019}. Other studies estimated pressure drops from PC-MRI velocity data using work-energy methods \cite{Donati:2015}, virtual work-energy approaches \cite{Marlevi:2020}, or turbulence-based quantities such as turbulent kinetic energy and turbulence production to estimate irreversible pressure loss in stenotic flows \cite{Dyverfeldt:2013,Casas:2016,Ha:2017}. More recently, data-driven approaches have been proposed to estimate pressure drop directly from 4D-flow MRI velocities in stenotic phantoms and patient data \cite{Nath:2023}. These methods can provide pressure information throughout the flow field, but they require time-resolved volumetric velocity data and additional post-processing.

% Quantification of turbulent energy losses
The need to represent pressure-loss mechanisms more explicitly is also supported by studies that quantify turbulence-related energy losses. Turbulent losses and pressure recovery are not fully accounted for in the SB and EB models. Quantification of turbulence has emerged as a promising avenue to improve the estimation of trans-stenotic pressure drop. Ha et al. \cite{Ha:2017} demonstrated that turbulence production defined using 4D-flow MRI correlates strongly with irreversible pressure-losses across stenoses. Similarly, Casas et al. \cite{Casas:2016} showed that quantification of turbulent kinetic energy (TKE) can provide geometry-dependent estimates of net trans-stenotic pressure drops, with more severe stenoses exhibiting higher turbulent energy dissipation. This finding further supports the necessity to account for turbulent energy dissipation for an accurate estimation of lesion severity \cite{Dyverfeldt:2013}, especially during exercise or pharmacologic stress.

% Scientific gaps
This link to turbulent energy dissipation highlights that the pressure-loss coefficient depends on the flow regime, which is typically characterized by the $Re$. However, many clinically established estimation methods still assume a constant flow regime with independent correlations. This flow regime sensitivity has profound clinical implications, particularly for quantifying stenosis severity. To illustrate: a stenosis assessed as moderate at rest may behave quite differently during exercise when flow rates increase and the flow becomes turbulent\cite{Marechaux:2010,Otto:1992,Gorlin:1951}. Several studies have used analytical solutions for flow through constricted geometries to incorporate flow regime effects into the model \cite{Young:1973,Morgan:1974,Young:1973b,Padmanabhan:1980}. Young and Tsai \cite{Young:1973} indicated that pressure-loss mechanisms vary significantly across laminar, transitional, and turbulent regimes. For steady flow they showed that the pressure drop can be written as the sum of a viscous term and a turbulence dissipation term. This approach was later extended to account for the effects of
pulsatile flow \cite{Young:1973b} and stenosis shape \cite{Seeley:1976}. However, their correlation, expressed solely in terms of $Re$, cannot be directly applied in routine clinical settings particularly with Doppler echocardiography and PC-MRI, which primarily provide peak velocity and flow rate. A recent study derived the SB expression in terms of flow rate for cerebral venous flow and reported good agreement with CFD \cite{Sidora:2025}.This indicates that deriving an SB model in terms of flow rate can work in their specific application. However, this approach does not account for turbulence-related losses, which is acceptable in cerebral venous flow where this can likely be ignored.

In summary, the image-based approaches discussed above are valuable because they can estimate relative pressure distributions or pressure-loss fields from spatially and temporally resolved velocity data. However, in many clinical settings, the main quantity of interest is the overall trans-stenotic pressure drop rather than the full pressure field \cite{Baumgartner:2017}. For this purpose, Bernoulli models remain attractive because they are simple, fast, and can be applied using reduced velocity quantities such as flow rate, bulk velocity, or peak stenotic velocity. However, conventional Bernoulli formulations do not explicitly account for the $Re$ dependence of viscous, turbulent, and pressure-recovery losses. 
% Aim
The present study therefore focuses on improving these Bernoulli models by introducing a modified Bernoulli formulation in which the pressure-loss coefficient varies with $Re$, so that the global pressure drop can account more explicitly for changes in viscous, inertial, turbulence-related, and pressure-recovery losses.  In this way, the model retains the practical advantage of using the bulk or peak velocity only, while incorporating flow-regime-dependent loss behavior that is absent from the simplified and extended Bernoulli equations. To evaluate the model, velocity fields were measured using ultrasound imaging velocimetry (UIV) and PC-MRI, together with direct pressure-drop measurements. In addition, when velocity information is obtained using MRI, the model accuracy depends strongly on in-plane pixel size, which is addressed in the second part of the present study. By examining a range of $Re$ numbers spanning laminar and turbulent regimes, we aim to: (1) establish the relationship between peak and bulk velocities in the stenosis throat, the associated $Re$-dependent loss mechanisms, and pressure drop, and (2) integrate MRI-based velocities into this framework to quantify how a pixel-size--dependent velocity bias affects the trans-stenotic pressure drop predictions.

\section*{Methods}

\subsection*{Bernoulli Methods for pressure drop Estimation}
\subsubsection*{Simplified Bernoulli Equation}
The SB equation represents the most clinically prevalent method for pressure estimation \cite{Ha:2017}. This method is derived from the Navier-Stokes equation under several simplifying assumptions: inviscid flow, negligible body forces, steady flow conditions, and negligible upstream velocity. Under these assumptions, the pressure difference is given by:

\begin{equation}
    \Delta p_{\text{SB}} = p_1 - p_2 \approx \frac{1}{2}\rho V_\text{peak,t}^{2}
    \label{eq:psb}
\end{equation}
where $p_\text{1}$ and $p_\text{2}$ are the upstream and downstream pressures respectively, and $V_\text{ {peak,t}}$ is the peak velocity at the stenosis. In clinical applications, especially echocardiography, the pressure drop in mmHg is reported as \cite{Zardkoohi:2019}:

\begin{equation}
    \Delta p_{\text{SB}} = p_1 - p_2 \approx 4V_\text{peak,t}^{2}
    \label{eq:psbmmhg}
\end{equation}
assuming a constant blood density $\rho_ {b} =  {1060}$ kg/m$^3$ and using 1 mmHg $ = $ 133.322 Pa for unit conversion.

The pressure-loss coefficient for the SB method is defined as the ratio of the measured pressure drop and the dynamic pressure:

\begin{equation}\label{eq:Ksb}
    K_\text{SB} = \frac{\Delta p_\text{SB}}{\frac{1}{2}\rho V_\text{bulk}^2} = \frac{ V_\text{peak,t}^2}{ V_\text{bulk}^2}
\end{equation}
where $V_\text{ {bulk}}$ is the bulk velocity in the upstream vessel. This dimensionless pressure-loss coefficient enables comparison for different geometries and operating conditions.

\subsubsection*{Extended Bernoulli Equation}
The EB equation accounts for the geometric properties of the stenosis through the effective orifice area and the anatomical cross-sectional area, resulting in an area-ratio correction that captures potential downstream pressure recovery \cite{Garcia2000}.  

\begin{equation}
    \Delta p_{\text{EB}} = p_1 - p_2  \approx \frac{1}{2}\rho V_\text{peak,t}^{2} \left(1-\frac{A_\text{t}}{A_\text{A}}\right)^2 =  \Delta p_{\text{SB}} \left(1-\frac{A_\text{t}}{A_\text{A}}\right)^2 \quad 
    \label{eq:peb}
\end{equation}
where $A_{\text{t}}$ is the effective orifice area, and $A_ \text{A}$ is the upstream anatomical cross-sectional area of the vessel. The correction factor $(1-\frac{A_\text{t}}{A_\text{A}})^2$ accounts for the geometric contraction and provides a more accurate representation of the pressure-velocity relationship in stenotic flows compared to the SB approach.

The pressure-loss coefficient for the EB method can be calculated as:

\begin{equation}\label{eq:Keb}
    K_\text{EB} = \frac{\Delta p_\text{EB}}{\frac{1}{2}\rho V_\text{bulk}^2} = \frac{ V_\text{peak,t}^2}{ V_\text{bulk}^2} \left(1-\frac{A_\text{t}}{A_\text{A}}\right)^2
\end{equation}
where $\Delta  {p}_ \text{EB}$ represents the pressure drop from the EB approach.

\subsubsection*{Modified Bernoulli Equation}
For steady flow, the pressure drop from the modified Bernoulli (MB) equation can be written as the sum of a viscous term and a turbulent term \cite{Young:1973}:

\begin{equation}
    \Delta p_\text{MB} = \frac{1}{2}\rho V_\text{bulk}^2 \left(\frac{k_\text{v}}{Re} + k_\text{t} \left(1-\frac{A_\text{A}}{A_\text{t}}\right)^2\right) = \frac{1}{2}\rho V_\text{bulk}^2 K_\text{MB}
\end{equation}
where the first term represents viscous losses, the constant $k_\text{v}$ is a geometry-dependent coefficient and $Re$ is the Reynolds number $ Re = \frac{\rho V_{bulk}D}{\mu}$  $=\frac{4\rho Q}{\pi \mu D}$, where $D$ is the inlet pipe diameter, $\mu$ is the dynamic viscosity, and $Q$ is the volumetric flow rate. The second term reflects losses due to turbulence, and the coefficient $k_\text{t}$ is the expansion loss coefficient. The corresponding pressure-loss coefficient is defined as:

\begin{equation}\label{eq:Kmb1}
    K_\text{MB} = \frac{\Delta p_\text{MB}}{\frac{1}{2}\rho V_\text{bulk}^2} = \frac{k_\text{v}}{Re} + k_\text{t} \left(1-\frac{A_\text{A}}{A_\text{t}}\right)^2  
\end{equation}

In this study, the coefficients $k_\text{t}$ and $k_\text{v}$ were obtained by fitting Eq. \eqref{eq:Kmb1} to the measured pressure drop data $\Delta {p}_\text{EXP}$. The bulk velocity $V_\text{bulk}$ was determined via UIV measurements upstream of the idealized model. The values obtained were $k_\text{v}=$  $62.4 \times 10^{3}$ and $k_\text{t}=$ 1.41. To assess the sensitivity of the calibrated coefficients to the selected $Re$ range, an internal hold-out and sensitivity analysis was also performed. In this analysis, $k_\text{v}$ and $k_\text{t}$ were refitted after excluding different regions of the calibration dataset. The tested cases included fitting to the full dataset, fitting only the low-$Re$ data, fitting only the high-$Re$ data, and fitting after omitting an intermediate $Re$ range. The resulting coefficients were then used to reconstruct K($Re$) and to calculate MB pressure drops from UIV-derived peak throat velocities. This analysis was used to evaluate the robustness of the fitted formulation within the available phantom dataset, but should not be interpreted as independent validation. Further details are provided in the Supplementary Material.

The Re-dependent pressure-loss coefficient captured the transition from high loss coefficients at low $Re$ to an asymptotic value at high $Re$. Using the definition of $Re$, we define $K_ \text{MB}$ in terms of flow rate through the relation:

\begin{equation}\label{eq:Kexpq}
    K_\text{MB} = \frac{k_\text{v}\pi \mu D}{4\rho Q} + k_\text{t}\left(1-\frac{A_\text{A}}{A_\text{t}}\right)^2 = \frac{k'_\text{v}}{Q} + k_\text{t} \left(1-\frac{A_\text{A}}{A_\text{t}}\right)^2 
\end{equation}
with ${k'_\text{v}}=\frac{k_\text{v}\pi \mu D}{4\rho}$. Also, by defining $C$ as the velocity ratio at the throat, $C =$ $V_\text{ {peak,t}}$/${V_\text{bulk,t}}$, where $V_ \text{{bulk,t}}$ is the bulk velocity in the throat, the pressure-loss coefficient using the MB can be defined based on $V_\text{peak,t}$: 

\begin{equation}\label{eq:Kmb}
    K_\text{MB} = C\frac{k'_\text{v}}{A_\text{t}V_\text{peak,t}} + k_\text{t} \left(1-\frac{A_\text{A}}{A_\text{t}}\right)^2 = \frac{k''_\text{v}}{V_\text{peak,t}} + k_\text{t} \left(1-\frac{A_\text{A}}{A_\text{t}}\right)^2
\end{equation}
where ${k''_\text{v}}=C\frac{k'_\text{v}}{A_\text{t}} = C\frac{k_\text{v}\pi \mu D}{4\rho A_\text{t}}$. Using the SB formulation of the pressure drop (Eq. \ref{eq:psb}), the pressure drop obtained using the MB equation becomes:

\begin{equation}
    \Delta p_\text{MB} = \frac{1}{2}\rho\left(\frac{Q}{A_\text{A}}\right)^2  K_\text{MB} = \frac{1}{2}\rho\left(\frac{V_\text{bulk,t}A_\text{t}}{A_\text{A}}\right)^2 K_\text{MB} = \Delta p_\text{SB}\left(\frac{A_\text{t}}{A_\text{A}}\right)^2\frac{ K_\text{MB}}{C^2} 
    \label{eq:pmb}
\end{equation}
Importantly, $K_\text{MB}$ in this expression is not a constant but depends on $V_\text{peak,t}$ as shown in Eq. \ref{eq:Kmb}.

\subsection*{\textit{In‑vitro} experimental facility}
% Experimental facility
The experiments are performed in a closed-loop facility, shown in Figure~\ref{fig:pipeline}a. The flow is driven by a centrifugal pump (compactON 2100, EHEIM GmbH, Deizisau, Germany) that recirculates water from a reservoir. The flow rate was adjusted with a manual valve that is installed after the pump. The flowrate and water temperature were monitored using an electromagnetic in-line flowmeter (AF-E 400, KROHNE, Duisburg, Germany), that is located after the valve. Subsequently, there is a 6 m straight polymethyl methacrylate (PMMA) pipe with inner diameter D $=$ 20 mm and the Food and Drug Administration (FDA) nozzle geometry. A flexible hose closes the flow loop, connecting the PMMA pipe to the reservoir. During the experiments the water is kept at a temperature of approximately T$ = 22^{\circ}C$ to ensure constant fluid properties ($\rho \approx 998$ kg/m $^3$, $\mu \approx 1.0 \times 10^{-3}$ Pa·s). Pressure measurements were obtained using Validyne DP15 variable reluctance pressure sensor (Validyne Engineering Corporation, Northridge, CA, USA), which were placed at two locations P$_ \text{1}$ and P$_ \text{2}$ to capture pressure variations across the test section (Fig. \ref{fig:pipeline}a). Data were acquired using a National Instruments DAQ (Austin, TX, USA), recorded at 1000 Hz for 10 s for each measurement. The experimental flow rates, corresponding $Re$ number, and bulk and peak velocities obtained from the UIV analysis are reported in Supplementary Table~S1. 

% FDA Nozzle details
The experimental model was a transparent FDA benchmark nozzle milled from polymethyl methacrylate (PMMA; Perspex), and is shown in Fig.~\ref{fig:pipeline}b. The geometry was scaled-up by a factor of three with respect to the reference geometry \cite{Hariharan:2011dg}. This results in an overall length of L $=$ 104.5 mm, inlet and outlet diameters D $=$ 20 mm, and a conical convergence with a $10^{\circ}$ half-angle resulting in a throat diameter of D$_{\text{t}} = $ 6.6 mm and throat length of L$_\text{z} = $ 40 mm. To provide a fully developed inlet condition, the nozzle inlet was positioned 130 pipe diameters (130D) downstream of the flowmeter, resulting stable, reproducible velocity profiles at the measurement location. This geometry represents a canonical case for an idealized stenosis model \cite{Hariharan:2011dg}.

\subsection*{UIV acquisition and processing}
Images were acquired with a Verasonics Vantage research ultrasound system (Verasonics Inc., Kirkland, WA, USA). The system is coupled to a linear array transducer (L11-5v) of 128 elements with a pitch of 0.3 mm and an elevation focus at 18 mm. The probe was operated at a center frequency of 7.5 MHz (bandwidth 6.5-8.5 MHz) and provided a lateral field of view of L$_\text{{z}} =$ 38.4 mm along the array and a depth of 30 mm. Image acquisition was performed using parallel beamforming, with all array elements transmitting and receiving simultaneously. The measurement geometry and axes are shown in Fig.~\ref{fig:pipeline}c: the imaging depth $y$ was measured downward from the transducer surface into the flow. To improve the signal-to-noise (SNR) ratio of the ultrasound images, the PMMA wall thickness at the location of the throat was reduced from 16.7 mm to 4.4 mm \cite{Ono:2020,Gurung:2016,Poelma:2017}. The space between the flat surface created after removing the wall and the transducer was filled with water to provide acoustic coupling. The flow was seeded with Vestosint particles (Degussa-Hüls, Frankfurt, Germany) with a mean diameter d$_\text{p} = {56}$ $\mu$m and density $\rho_\text{p} = {1016}$ g/cm$^3$.

The acquisition rate was varied from 600 to 1600 Hz, dependent on the flow velocity. For each flow condition, 2000 frames were recorded, resulting in 1000 image pairs for correlation. The grayscale image in Fig.~\ref{fig:pipeline}d (left) shows the B-mode images, while the right image shows the background subtracted B-mode image; red arrows indicate the local velocity vectors. The B-mode images were acquired using a standard single-angle plane-wave flash sequence. Velocity fields were then obtained using a modified version of PIVware MATLAB script \cite{Westerweel:1993}. We perform two-pass processing without any advanced techniques such as window offsets or iterative image deformation \cite{Westerweel:1997,Scarano:2001}.  Interrogation windows of [16 64] pixels followed by [8 32] pixels were used, each with 50$\%$ overlap. Correlation averaging was applied by averaging the cross-correlation maps over the full set of image pairs before estimating the final velocity vectors.

Following the resolution estimates described by Poelma \cite{Poelma:2017}, the axial ultrasound resolution for pulsed imaging is expected to be on the order of the acoustic wavelength, while the lateral resolution depends on aperture, focal length, wavelength, and beamforming. The ultrasound imaging resolution was measured experimentally using a 0.12 mm wire target in water with the same sequence and acquisition parameters as the UIV measurements. The measured full width at half maximum (FWHM) was $0.303 \pm 0.027$ mm in the axial direction and $0.793 \pm 0.068$ mm in the lateral direction. Further details on the wire-target measurement, ultrasound sampling, acoustic length scale, and UIV processing scale are provided in the Supplementary Material.

Prior to the pressure drop analysis, the accuracy of the UIV measurements was validated through comparison with theoretical bulk velocity calculations. Velocity measurements were conducted upstream of the idealized stenosis model in the straight pipe section and in the throat section for flow rates ranging from 1 to 4 L/min. The bulk velocity from UIV measurements $V_{\text{bulk}}$ was calculated using radial integration, assuming a symmetric profile which was confirmed by the velocity profile:
\begin{equation}
    {V}_{{bulk}} = \frac{2}{R^2}\int_0^R u(y) \, y \, dy
    \label{eq
    :bulk_velocity_UIV}
\end{equation}
where $u(y)$ represented the measured velocity profile as a function of radial position $y$ according to Fig.~\ref{fig:pipeline}a.

\subsection*{MRI acquisition and processing}
MRI measurements were obtained with a 3 T preclinical system (MR Solutions, UK); using a 2D gradient-echo phase-contrast sequence \cite{Bernstein:2004MRI}.  Symmetric velocity encoding was used, in which equal and opposite encodes ($+V_{\text{enc}}$ and $-V_{\text{enc}}$) were acquired and differenced, rather than using a separate flow-compensated reference. Further sequence-design details are provided in the Supplementary Material.
The MRI signals were acquired using a quadrature receiver coil with a diameter of 54 mm. The center of a single slice (thickness 5 mm) was positioned at the minimum cross-section at $z =$ -0.036 m with $z =$ 0 at the beginning of the expansion (Fig.~\ref{fig:pipeline}f). Although we report in-plane pixel size, velocities are voxel-averaged because the slice thickness is 5 mm. For each flow rate in the range 1.5 to 3.5 L/min, one dataset was acquired at an in-plane pixel size of 0.13 mm/px, and the coarser pixel sizes (0.25, 0.50, 1.00, and 1.33 mm/px) were generated retrospectively during reconstruction; further details are provided in the Supplementary Material. MRI bulk velocities $V_ \text{{bulk}}$ were obtained by integrating over the segmented lumen, compared to bulk velocity from the flowmeter, and the relative error was computed for every flow-rate--pixel-size pair. Table~\ref{tab:mriacq} shows the parameters that were used for the MRI acquisitions and the reconstruction matrix sizes.

\begin{table}[ht]
        \centering
        \begin{tabular}{l l l l}
        \hline
        Parameter &  &  & Value \\
        \hline
         Field of view (FOV)              &   &  & 32 mm $\times$ 32  mm \\
         Acquired matrix size (read $\times$ phase $\times$ slice)                                                         &   &  & 256 $\times$ 256 $\times$ 1 \\
         Acquired nominal in-plane pixel size 
         &   &  & 0.13 mm $\times$ 0.13 mm \\
         
        % Acquired nominal voxel size 
        % &   &  & 0.13 mm $\times$ 0.13 mm $\times$ 5 mm \\
         Acquired in-plane FWHM (read $\times$ phase)
          &   &  & 0.238 mm $\times$ 0.252 mm  \\
         Reconstructed matrix sizes (read $\times$ phase $\times$ slice)                                                   &   &  & 128 $\times$ 128 $\times$ 1
                                         , 64 $\times$ 64 $\times$ 1
                                         , 32 $\times$ 32 $\times$ 1
                                         , 24 $\times$ 24 $\times$ 1 \\

        Reconstructed nominal in-plane pixel sizes 
                                         &   &  &  0.25 mm, 0.50 mm, 1.00 mm, 1.33 mm \\
        Slice thickness                  &   &  & 5 mm  \\
        Repetition time (TR)            &   &  & 50 ms \\
        Echo time (TE)                  &   &  & 3 ms \\
        RF flip angle ($\alpha$)         &   &  & 10$^{\circ}$  \\
        Velocity encoding ($V_\text{enc}$)       &   &  & 110-300 cm/s  \\
        Number of averages              &   &  & 10  \\
        %Acquisition time including 10 averages                &   &  &  4.3 min per acquired dataset  \\
        Time per average
        &   &  & 25.8 s \\
        \hline
        \end{tabular}
        \caption{\label{tab:mriacq} MRI parameters used.}
\end{table}

The nominal in-plane pixel size was calculated from the field of view (FOV) and matrix size; however, this value represents the reconstructed sampling grid and should not be interpreted as the effective spatial resolution. The effective MRI in-plane spatial resolution was estimated using an edge-spread-function (ESF) approach, following the MRI resolution assessment framework described by Delakis et al. \cite{Delakis:2009}. Briefly, a static water-filled PMMA pipe was scanned without flow using the same imaging parameters as the velocity measurements, and the water--PMMA boundary was used as a high-contrast edge. The measured MRI FWHM was 0.238 mm in the read direction and 0.252 mm in the phase direction, compared with a nominal in-plane pixel size of 0.125 mm, reported as 0.13 mm. Further details of the ESF analysis are provided in the Supplementary Material.

\subsection*{Calibrating and applying the MB model}
\begin{figure}[ht!]
    \centering
    \includegraphics[width=\linewidth]{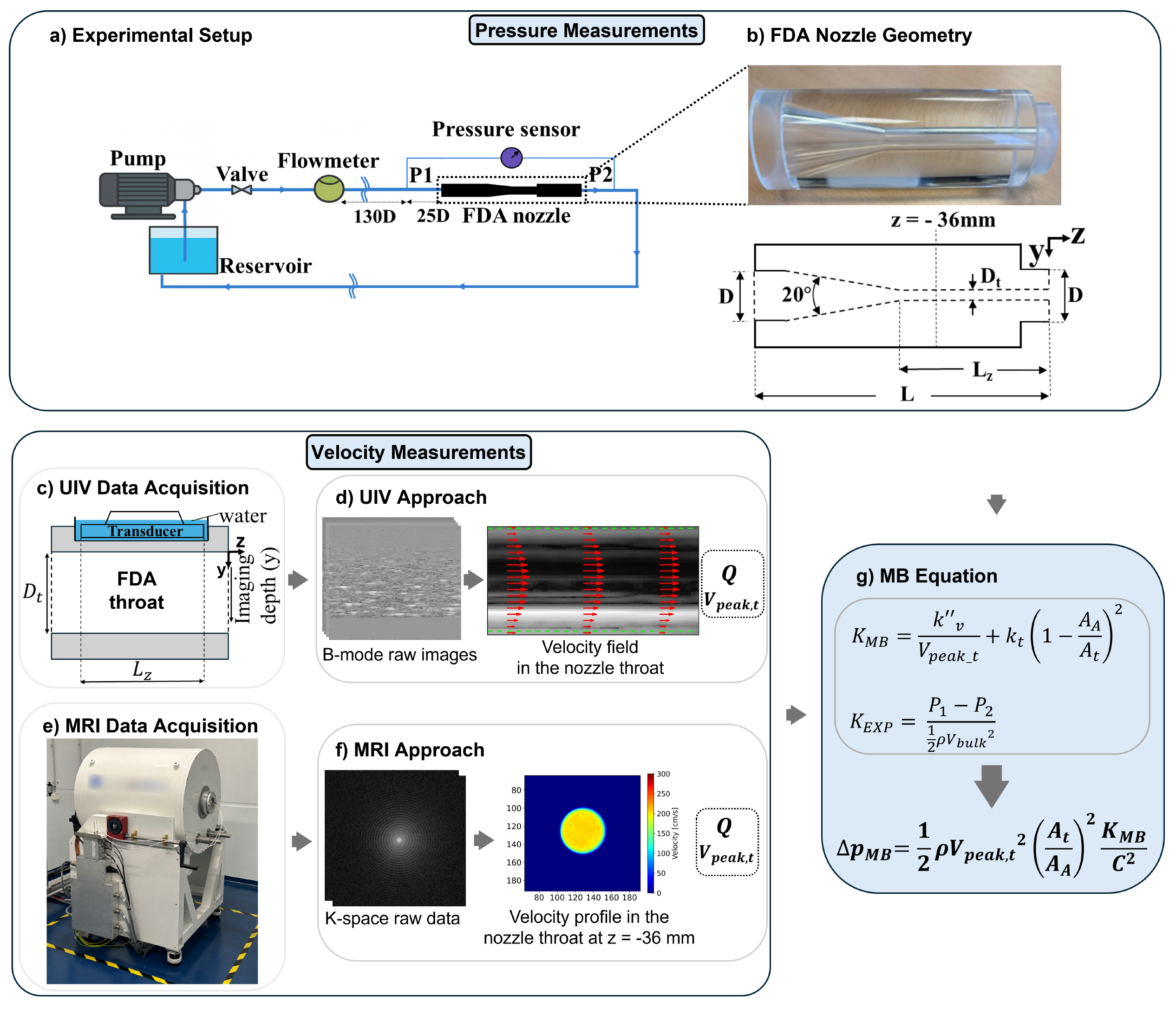}
    \caption{Overview of experimental approach and methodology: (a) closed-loop flow rig; (b) the FDA nozzle geometry and dimensions; (c) ultrasound imaging configuration at the nozzle throat; (d) UIV approach; (e) image of the MRI system and part of the experimental facility; (f) MRI approach; (g) MB formulation, where the pressure-loss coefficient $K_\text{MB}$ is calibrated using the experimental coefficient $K_\text{EXP}$.} 
    \label{fig:pipeline}
\end{figure}

Figure~\ref{fig:pipeline} summarizes the experimental and analytical workflow used to define the MB model. Pressure drops across the idealized stenosis model were measured between pressure taps P$_\text{1}$ and P$_\text{2}$ for $Re$ corresponding to flow rates of $0.65-3.9$ L/min (Fig.~\ref{fig:pipeline}a), and time-averaged to obtain the experimental pressure-loss coefficient and calibrate $K_\text{MB}$ (Eq.~\eqref{eq:Kmb}), yielding $k''_\text{v}$ (or $k'_\text{v}$) and $k_\text{t}$. Peak velocities, the throat velocity ratio, and flow rate were then measured with UIV and PC-MRI and used as inputs to the SB, EB, and MB formulations for $Q=0.85 - 3.5 $ L/min (Figs.~\ref{fig:pipeline}c-f), enabling comparison of model predictions against the measurements (Fig.~\ref{fig:pipeline}g). The error analysis for the pressure-loss coefficient, pressure drop, and bulk velocity was based on the percentage error, root-mean-square error (RMSE), and mean absolute percentage error (MAPE), defined as follows:
%The errors for the pressure-loss coefficient were defined as follows:

\begin{equation}
     {\text{Percentage Error}} = \frac{ {K}_{ {MODEL},i} -  {K}_{ {EXP},i}}{ {K}_{ {EXP},i}} \times 100%
    \label{eq:percentage_error}
\end{equation}

\begin{equation}
     \text{RMSE} = \sqrt{\frac{1}{ {N}}\sum_{i=1}^{ {N}}\bigl( {K}_{ {MODEL},i}- {K}_{ {EXP},i}\bigr)^{2}}
    \label{eq:rmse}
\end{equation}

\begin{equation}
      \text{MAPE} = {100}\times\frac {1}{ {N}}\sum_{i=1}^{ {N}}\left|\frac{ {K}_{ {MODEL},i}- {K}_{ {EXP},i}}{ {K}_{ {EXP},i}}\right|
    \label{eq:mape}
\end{equation}
where $i$ indexes the measurement cases (e.g., each flow rate condition) and $N$ is the total number of cases included in the error calculation.

\section*{Results}

\subsection*{Pressure-loss coefficient}
The pressure-loss coefficient over the stenotic geometry as function of $Re$ is shown in Fig.~\ref{fig:p}a. The experimental values (EXP, black triangles) show a significant dependence on $Re$, decreasing from roughly 180 at $Re=$  1000 to about 107 at $Re=$  4000. Additionally, the predicted pressure-loss coefficient by the different models (MB, SB, and EB) are added. A fitted empirical correlation (FIT, solid black line) based on the Young relation \cite{Young:1973,Young:1973b} reproduces this trend well, showing higher losses at low $Re$ and approaching an asymptotic value as $Re$ increases. The UIV-derived parameters $V_{\text{peak,t}}$ and $C$ were measured for six flow rates and listed in Supplementary Table~S1. For the MB model (Eq. \ref{eq:Kmb}), $V_{\text{peak,t}}$ and $C$ were used as inputs. The results from the SB and EB models were obtained using only $V_{\text{peak,t}}$ as input. The MB model with $C_{\text{mean}}=1.42$ (green symbols) follows the experimental data closely over the entire $Re$ range, with a maximum relative error of 5$\%$  (Fig.~\ref{fig:p}a). The gray band around it shows the effect of varying $C$ between 1.37 and 1.47, indicating that the MB predictions are only weakly sensitive to the uncertainty in $C$ and remain close to the measurements across the tested range. In contrast, the SB model (blue circles) overpredicts the pressure-loss coefficient, especially for $Re$ above 2000 with relative errors of 10 to 55$\%$. The EB model (red squares) underpredicts the pressure-loss coefficient as $Re$ increases up to about 2300, and then overestimates the pressure-loss coefficient at higher $Re$ above 3000, with errors spanning roughly -15 to +25$\%$. Overall, the MB model provides the best agreement with the measured pressure-loss coefficients for the idealized stenosis model.

\subsection*{UIV-based trans-stenotic pressure drop estimation}

\begin{figure}[t]
      \centering
      \begin{tikzpicture}
        \node[inner sep=0] (img) {\includegraphics[width=\linewidth]{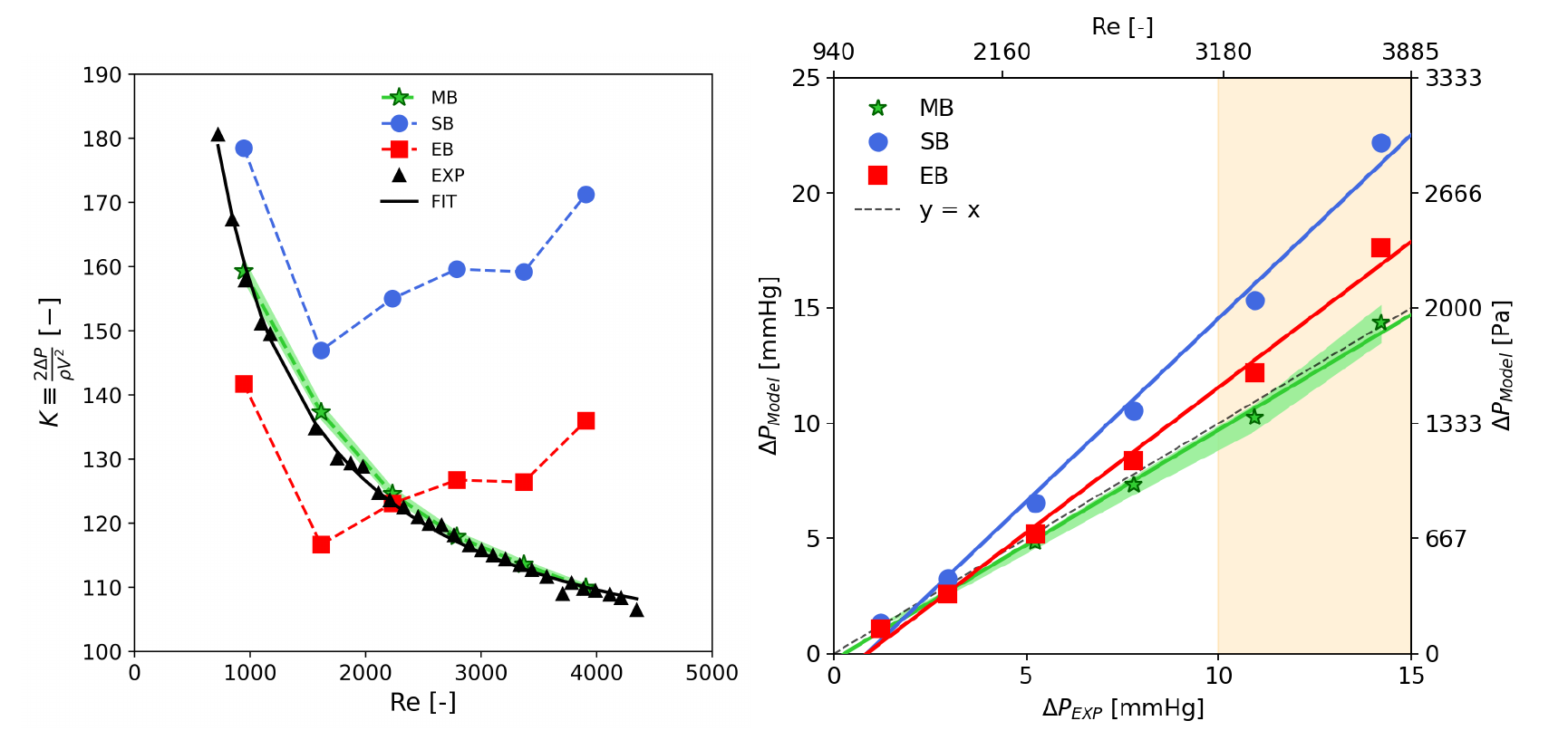}};
        % a) at top-left
        \node[anchor=south west, font=\bfseries] at (img.north west) {a)};
        % b) shifted to the right
        \node[anchor=south west, font=\bfseries, xshift=0.5\linewidth] at (img.north west) {b)};
      \end{tikzpicture}
    
      \caption{(a) Pressure-loss coefficient, $K$ as function of $Re$. (b) Predicted pressure drop, $\Delta P_{{Model}}$, versus measured trans-stenotic pressure drop, $\Delta P_{{EXP}}$. MB (green; dashed) uses $C_\text{{mean}}$; the gray band shows the range for $C=$ 1.37-1.47. SB = simplified Bernoulli; EB = extended Bernoulli; FIT = fitted correlation of experimental data (EXP). Pressure is reported in mmHg (bottom/left) and Pa. The dotted line in (b) is the identity line $y=x$. The yellow shaded band indicates clinically relevant $Re$ number range for stenotic flows.}
      \label{fig:p}
\end{figure}

Figure~\ref{fig:p}b compares the measured pressure drop across the idealized stenosis model with predictions from the SB, EB, and MB models, where the black dashed $y=x$ line indicates perfect agreement. The yellow shaded band highlights the clinically relevant range in this model, corresponding to $Re\approx$ $2800-3900$ (i.e., $\Delta p=$ $10-15$ mmHg), where accurate pressure drop prediction is most important for clinical decision-making. Overall, the MB model with $C_\text{mean}=1.42$ matches the measurements closely over the full range ($1 - 15$ mmHg) with an average error of $5\%$ and a maximum error below 10$\%$. The gray MB band obtained by varying $C$ indicates that the predictions are weakly dependent on the uncertainty in $C$. In contrast, the SB model systematically overpredicts the pressure drop, with relative errors exceeding 50$\%$ and the bias increasing at higher values (up to about 22 mmHg predicted versus 14 mmHg measured). The EB model performs comparably to MB at low pressure drops, but starts overestimating for pressure drops above 6 mmHg with relative errors spanning from about -15 to +25$\%$, resulting in an intermediate performance between MB and SB at higher pressure drops. In the clinically relevant regime, the MB model performs well with a maximum error below 5$\%$. In contrast, the SB and EB models, overpredict the pressure drop with a maximum error of +55 and +25$\%$, respectively.

These trends are confirmed by a statistical analysis, which is presented in Table~\ref{tab:error}. We have included an ordinary least-squares regression, root mean square error (RMSE), and the mean absolute percentage error, MAPE, for each model. The SB consistently overestimates the pressure drop, with a regression slope of 1.59 (R${^2}=$ 0.992), large scatter (RMSE $=$ 3.93 mmHg) and the highest relative error (MAPE $=$ 29.9$\%$). The EB also overpredicts, but more moderately (slope 1.26, R${^2}=$  0.992, RMSE $=$ 1.51 mmHg, MAPE $=$ 11.1$\%$). As expected from the results in Fig.~\ref{fig:p}b, the MB approach using $C_\text{mean}$ follows the agreement line (dashed black line) almost one-to-one (slope 0.998, R${^2}=$ 0.996), with the smallest errors (RMSE $=$ 0.404 mmHg, MAPE $=$ 6.26$\%$). Overall, the MB model with a $Re$-dependent loss coefficient presents the lowest errors within the considered regime.

\begin{table}[ht]
    \centering
    \begin{tabular}{l l l l}
    \hline
    Method & MB at $C_\text{mean}$ & SB & EB \\
    \hline
    Regression slope & 0.998 & 1.59 & 1.26 \\
    
    R${^2}$ & 0.996  & 0.992 & 0.992 \\
    
    RMSE [mmHg] & 0.404  & 3.93 & 1.51 \\
    
    MAPE $\%$ & 6.26  & 29.9 & 11.1  \\
    \hline
    \end{tabular}
    \caption{\label{tab:error}Statistical comparison of model performance for predicting trans-stenotic pressure drops in the idealized stenosis model. Shown are the regression slope against the line of identity, coefficient of determination (R${^2}$), root mean square error (RMSE) and mean absolute percentage error (MAPE) for the MB, the SB, and the EB methods.}
\end{table}

\subsection*{PC-MRI velocity fields}
\begin{figure}[t]
      \centering
      \begin{tikzpicture}
        \node[inner sep=0] (img) {\includegraphics[width=\linewidth]{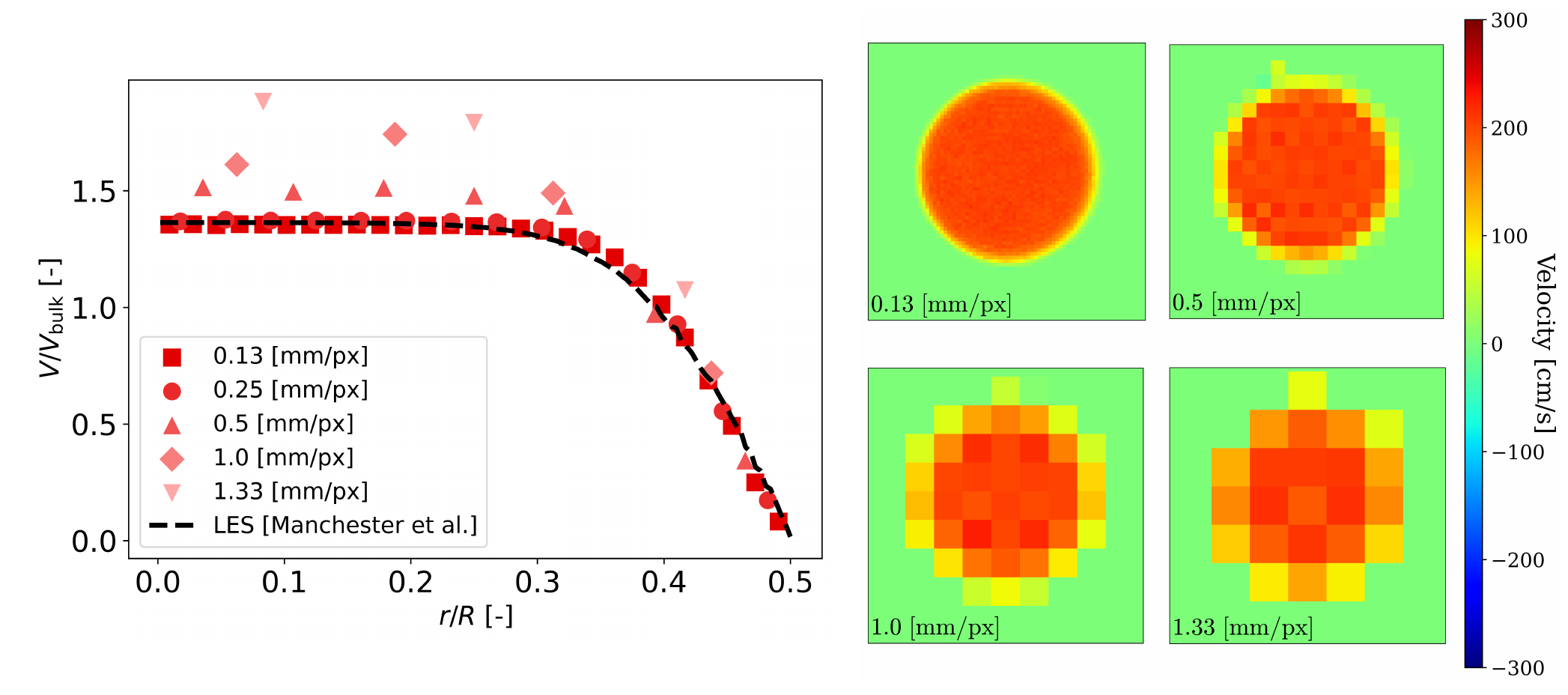}};
        % a) at top-left
        \node[anchor=south west, font=\bfseries] at (img.north west) {a)};
        % b) shifted to the right 
        \node[anchor=south west, font=\bfseries, xshift=0.5\linewidth] at (img.north west) {b)};
      \end{tikzpicture}
    
      \caption{Effect of in-plane pixel size on the masked velocity profile at $z = $ -0.036 m in the throat. (a) Normalized mean velocity profiles $V/V_\text{bulk}$ from PC-MRI for five different pixel sizes $0.13-1.33\ $ mm/px for $Re=$ 667, compared with the reference profile at the same $Re$ number \cite{Manchester:2020}. (b) Velocity maps at different pixel sizes, showing increasing PVE and a bias toward lower $V_{\text{bulk}}$ as pixel size increases.}
      \label{fig:mri1}
\end{figure}

Figure~\ref{fig:mri1}a shows the normalized mean velocity profiles at $Re=$ 667 for five different pixel sizes (0.13, 0.5, 1.0, and 1.33 mm/px) and a reference velocity profile obtained from large eddy simulation (LES) \cite{Manchester:2020}. For pixel sizes 0.13 and 0.25 mm/px, the lumen radius is sampled with 27 and 14 pixels, respectively, allowing the high-velocity core to be clearly resolved. At a pixel size of 0.50 mm/px, the lumen radius is sampled with 8 pixels, resulting in a more discretized velocity profile. This pixel density is still sufficient to capture the overall velocity profile. For the cases with a pixel size of 1.0 and 1.33 mm/px only 4 and 3 pixels represent the throat radius, respectively. This results in a velocity measurement that is dominated by partial volume effects (PVE), i.e., individual voxels contain a mixture of fast-moving fluid and slower moving flow, which smooths the velocity field and biases the voxel-averaged values \cite{Hamilton:1994}. Figure~\ref{fig:mri1}b illustrates representative PC-MRI velocity maps of the through-plane (axial) velocity component at the stenosis throat for four in-plane pixel sizes. It can be seen that the MRI data acquired with the smallest pixel size agrees well with the reference. However, increasing the pixel size leads to an underestimated mean velocity profile and smearing of the shear layer due to PVE.

\begin{figure}[t]
      \centering
      \begin{tikzpicture}
        \node[inner sep=0] (img) {\includegraphics[width=\linewidth]{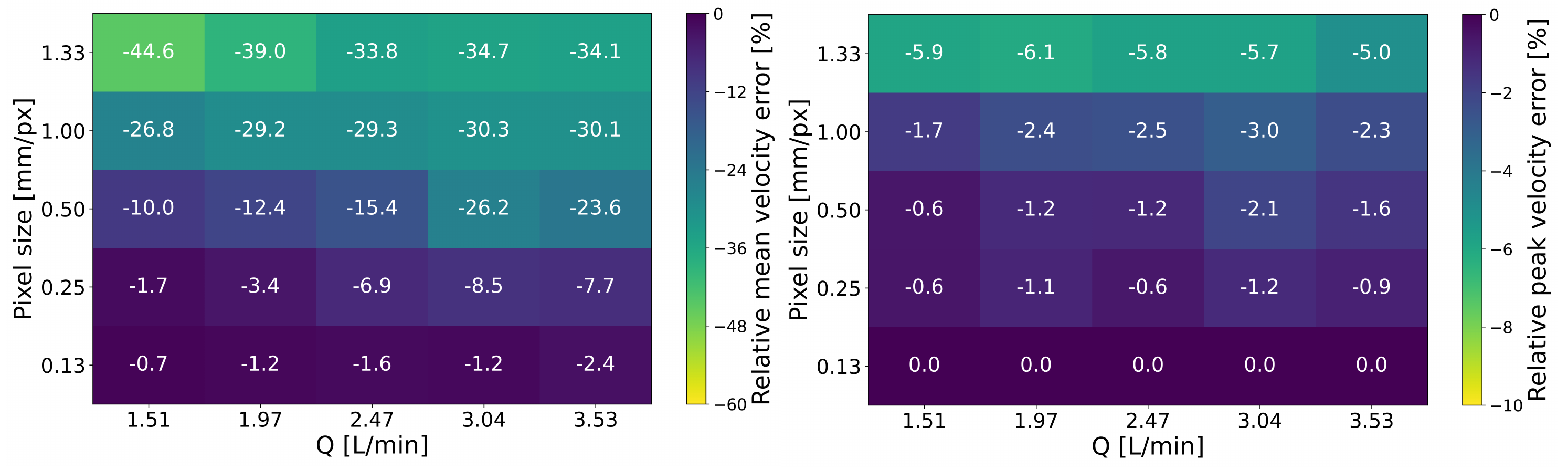}};
        % a) at top-left
        \node[anchor=south west, font=\bfseries] at (img.north west) {a)};
        % b) shifted to the right 
        \node[anchor=south west, font=\bfseries, xshift=0.5\linewidth] at (img.north west) {b)};
      \end{tikzpicture}
    
      \caption{Heatmaps showing the effect of in-plane pixel size on MRI-based throat velocities in the idealized stenosis model. (a) Percentage error in mean throat velocity versus pixel size and flow rate $Q$, with respect to the reference; negative values indicate underestimation. (b) Percentage error in peak throat velocity versus pixel size and flow rate $Q$, computed relative to the peak velocity measured at the smallest pixel size (0.13 mm/px).}
      \label{fig:mri4v}
\end{figure}

Figure~\ref{fig:mri4v}a summarizes the relative error in bulk velocity in the throat, computed by integrating the PC-MRI velocity over the segmented throat at z =  -0.036 and comparing against the reference bulk velocity from the flowmeter. At 0.13 mm/px, the bulk velocity bias remains small (approximately -0.7 to -2.4$\%$ across the studied flow rates), increasing to roughly -1.7 to -8.5$\%$ at 0.25 mm/px. At 0.50 mm/px the underestimation becomes substantial (about -10 to -26$\%$), and at $1.0 - 1.33$ mm/px the bulk velocity is underestimated by approximately -27 to -45$\%$, depending on the flow rate. Figure~\ref{fig:mri4v}b summarizes the relative error in the peak throat velocity, computed by comparing the peak velocity at each pixel size against the corresponding peak velocity obtained at the smallest pixel size (0.13 mm/px). In contrast to the bulk velocity, the peak velocity shows a significantly weaker dependence on pixel size: at 0.25 and 0.50 mm/px the differences remain small (typically within about -0.6 to -2.1$\%$), increasing to roughly -1.7 to -3.0$\%$ at 1.0 mm/px, and reaching approximately -5.0 to -6.1$\%$ at 1.33 mm/px, depending on the flow rate.

\begin{figure}[ht!]
    \centering
    \includegraphics[width=0.5\linewidth]{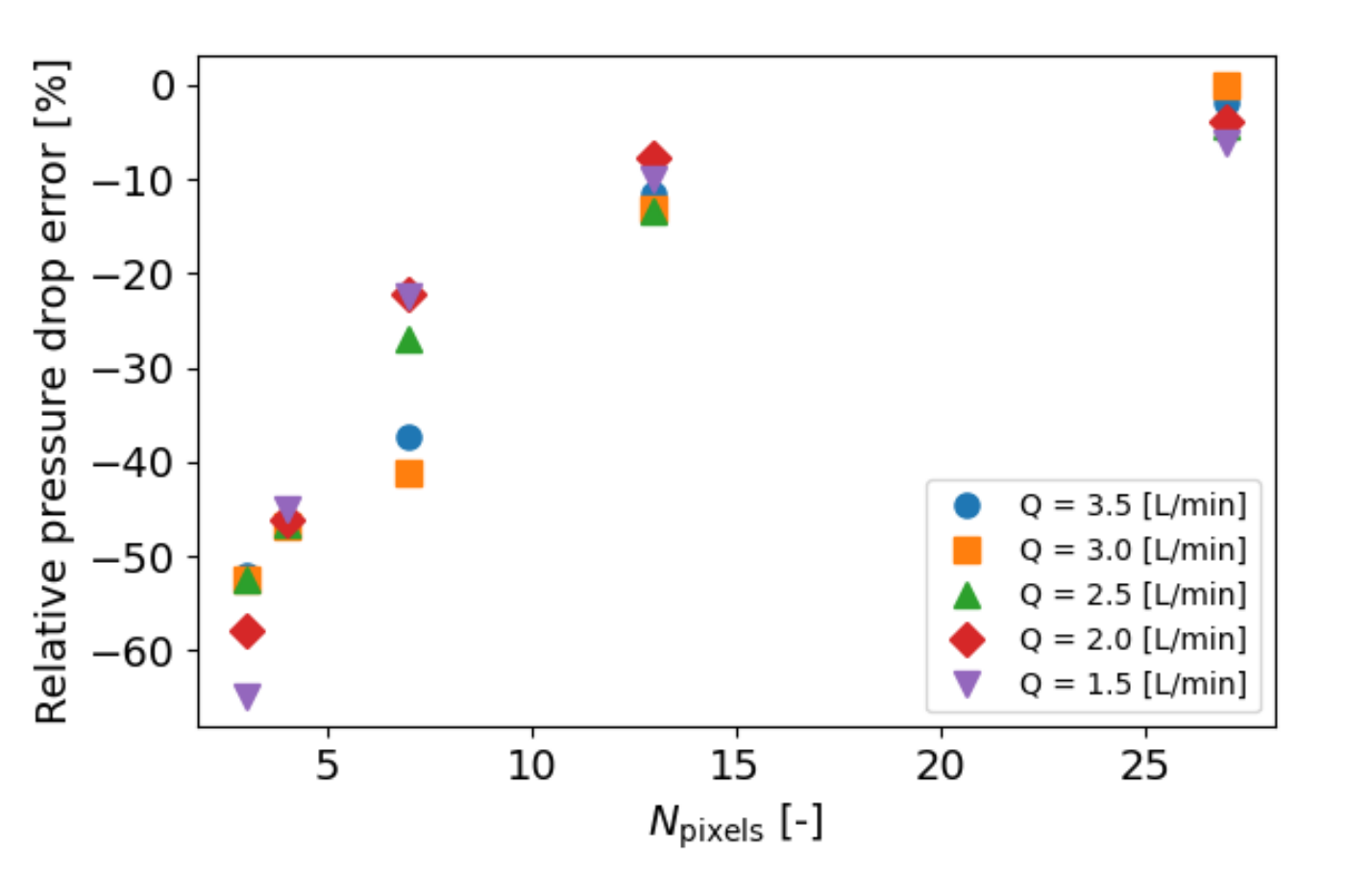}%
    \caption{The effect of the in-plane pixel size on MRI-based mean velocity in the throat of the idealized stenosis model. Percentage error in mean throat velocity versus the number of pixels across the throat radius ($N_{\text{pixels}}$) for five steady flow rates ($Q=1.5 - 3.5\ $ L/min) with respect to the reference flowrate measured by the flowmeter (negative indicates underestimation).}
    \label{fig:mri5}
\end{figure}

Figure~\ref{fig:mri5} represents the relative errors in the mean throat velocity in terms of the number of pixels across the throat radius, $N_{\text{pixels}}$. Across all flow rates, reducing the pixel size to only $3-4\ $ pixels across the throat radius ($N_\text{pixels}$) leads to substantial errors, with mean velocity underestimated by about -30 to -45$\%$. As $N_\text{pixels}$ increases, the error decreases monotonically. For $N_\text{pixels}$ $\approx$ 13, the velocity error falls below -10$\%$ for all flow rates. At the smallest pixel size with corresponding $N_\text{pixels}$ $\approx$ 27, the relative mean velocity error is reduced to a few percent only, ranging from -0.6 to -7$\%$.

\subsection*{MRI-based trans-stenotic pressure drop estimation}

\begin{figure}[t]
  \centering
  \begin{tikzpicture}
    \node[inner sep=0] (img) {\includegraphics[width=\linewidth]{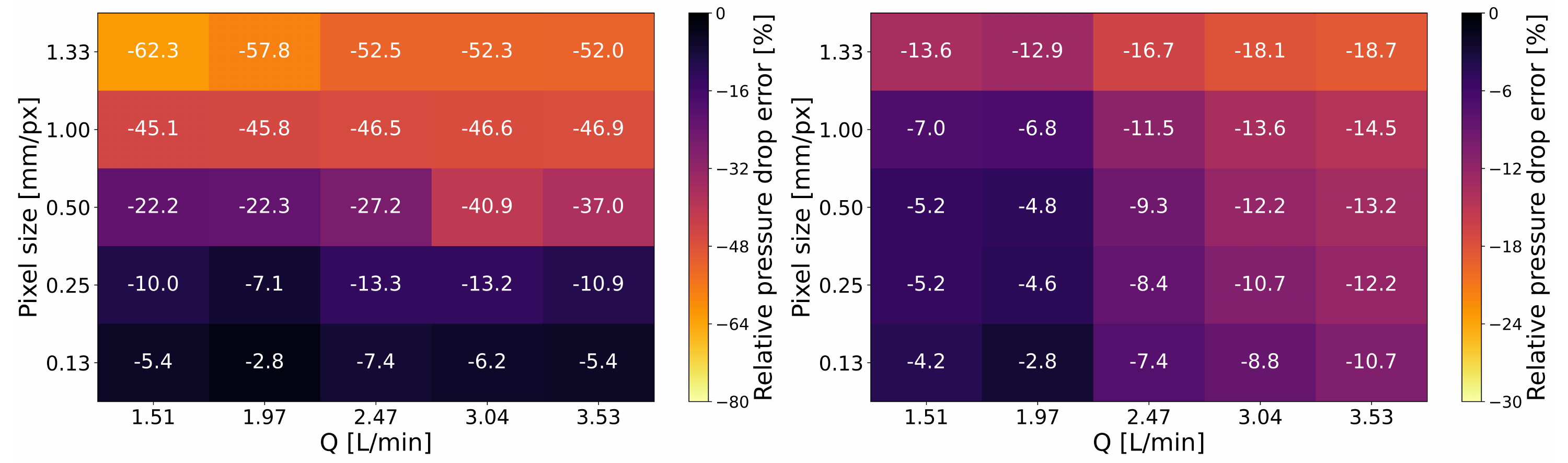}};
    \node[
      anchor=south west,
      font=\small,
      text width=0.45\linewidth,
      align=center,
      xshift=0.0\linewidth,
      yshift=0.6em
    ] at (img.north west) {\textbf{a)} Percentage error in MB-predicted pressure drop computed from the flow rate};
    \node[
      anchor=south west,
      font=\small,
      text width=0.48\linewidth,
      align=center,
      xshift=0.49\linewidth,
      yshift=0.6em
    ] at (img.north west) {\textbf{b)} Percentage error in MB-predicted pressure drop computed from the peak velocity};
  \end{tikzpicture}
  \caption{Heatmaps showing the effect of in-plane pixel size on MRI-based MB pressure drop estimates in the throat of the idealized stenosis model. (a) Percentage error in MB-predicted pressure drop computed from the flow-rate formulation (Eq. \ref{eq:Kexpq}) versus pixel size and flow rate $Q$, with respect to the measured pressure drop data; negative values indicate underestimation. (b) Percentage error in MB-predicted pressure drop computed from the peak-velocity formulation (Eq. \ref{eq:Kmb}) using $C_\text{mean}$ versus pixel size and flow rate $Q$, with respect to the measured pressure drop data.}
      \label{fig:mri4}
\end{figure}

The flow-rates derived from the PC-MRI data were used to estimate the trans-stenotic pressure drop using the MB formulation (Eq. \ref{eq:Kexpq}). These were compared against the measured pressure drops. As shown in Fig.~\ref{fig:mri4}a, the relative pressure drop error closely mirrors the bulk-velocity error. As the pixel size increases, the MB-predicted pressure drop is increasingly underestimated, ranging from about -6$\%$ at 0.13 mm/px (approaching -2$\%$ at the highest flow rates) to approximately -22 to -41$\%$ at 0.50 mm/px. At $1.0 - 1.33$ mm/px the underestimation reaches roughly -45 to -65$\%$. This indicates that, once the MB loss correlation is defined by fitting to the experimental pressure-loss coefficients, the dominant limitation in MRI-based pressure estimation is the pixel-size--dependent bias in the measured throat velocities rather than the contribution of the model itself. In Fig.~\ref{fig:mri4}b, the pressure drop is instead estimated using Eq. \eqref{eq:Kmb}, based on the peak throat velocity and $C_\text{mean}$ for all flow rates for the smallest pixel size (0.13 mm/px). In this case, the sensitivity to pixel size is markedly reduced: errors remain within approximately -3 to -13$\%$ across all flow rates, and increase only gradually toward the coarsest pixel sizes (up to about -18.7$\%$ at 1.33 mm/px). 

\section*{Discussion}

% Key points
In this study, we examined how different flow regimes influence the pressure-loss coefficient $K$ in an idealized stenosis model and introduced a Modified Bernoulli equation to predict the trans-stenotic pressure drop. The accuracy and systematic biases of various Bernoulli-based models were investigated by comparing them with accurate experimental reference data. Additionally, we evaluated how MRI in-plane pixel size affects velocity accuracy and the resulting pressure drop estimates using the proposed modified Bernoulli model.

% MB formulation
By incorporating flow physics into a pressure-loss coefficient formulation, we defined an MB model. The formulation of the MB model makes it flexible in terms of how it can be used in practice. Expressing the MB pressure drop in terms of the throat peak velocity $V_\text{peak,t}$ makes it directly compatible with Doppler echocardiography, which measures this quantity routinely. On the other hand, rewriting the MB equation in terms of flow rate removes the need for direct peak velocity measurements and allows the trans-stenotic pressure drop to be estimated from lumen area and flow, quantities that are accessible with PC-MRI or other imaging and flow-measurement techniques. 

Comparing SB, EB, and MB across the tested flow regimes, we found that the MB model provided the closest agreement with the experimental data. Even when using a single averaged value for peak to bulk velocity ratio ($C_\text{mean}=1.42$) rather than a case-specific ratio, the MB method kept errors within about $\pm 10\%$ across all flow conditions. In contrast, the SB model overestimated the pressure drop by up to 55$\%$, with errors increasing at higher $Re$, consistent with previous observations \cite{Kazemi2022,Ha:2017,Falahatpisheh:2015}. The EB model reduces this bias by accounting for the effective orifice area and associated pressure recovery, but still showed errors up to +25$\%$, demonstrating that a pressure-recovery correction alone does not fully compensate for the flow physics. 

% Re-dependent relation and turbulent flows
The key point is that pressure-loss is flow regime dependent \cite{White:2011}, and the relevant regime in a stenosis is set by the throat $Re$ number. In our approach, the throat $Re$ number is rewritten as $Re_\text{t}=(D/D_\text{t})\times Re$; for the present stenosis model $D/D_\text{t}=$ 3.03, giving $Re_\text{t}\approx$ $3000-12000\ $, indicating that turbulent effects are important in stenotic flows \cite{Hariharan:2011dg, Ahmed:1984,Manchester:2020,Varghese:2007}. Consistent with this, 4D-flow MRI studies have linked turbulence production and turbulent kinetic energy dissipation to the trans-stenotic pressure drop \cite{Ha:2017,Dyverfeldt:2013}. This explains why constant-coefficient SB and geometry-based EB formulations do not reliably capture losses as $Re$ increases. Along similar lines, Oshinski et al. \cite{Oshinski:1996} improved non-invasive gradients in aortic coarctation by using a stenosis-severity--dependent coefficient in $\Delta{{P}}=KV_{\text{peak,t}}^{2}$, but since $K$ does not explicitly depend on the $Re$ number, flow regime effects remain absorbed into a single empirical constant rather than as function of flow regime \cite{Oshinski:1996}. Therefore, it is important to take the flow regime into account when performing pressure measurements.

%kv and kt coefficients
In our approach, the pressure-loss coefficient $K$ is written as the sum of two contributions: a viscous term and a turbulent term. The fitted turbulent loss coefficient $k_\text{t}$ for the trans-stenotic gradient was 1.41, in close agreement with the commonly used value of 1.52 reported by Seeley and Young \cite{Seeley:1976}. In contrast, for the viscous resistance coefficient $k_\text{v}$, Seeley and Young showed that there was a stronger geometrical dependence: in our case a value of $k_\text{v} = 62.4 \times 10^{3}$ was found, whereas they reported values spanning $421$ to $56.2 \times 10^{3}$ (reported using the normalization $\Delta p/(\rho V^{2})$; expressed in our normalization $2\Delta p/(\rho V^{2})$ this corresponds to $842$ to $112.4 \times 10^{3}$). This wide span supports the idea that in particular $k_\text{v}$ is not a universal constant, but a parameter that should be reported and, when possible, tuned to a specific vessel or lesion. Recent work by Chernyavsky et al. \cite{Chernyavsky:2025} has shown that calibrating $k_\text{v}$ and $k_\text{t}$ to vessel elasticity and stenosis geometry substantially improves agreement between predicted and measured pressure drops, which consolidates our current observations.

% Discussion about C from UIV measurements
A critical component of the MB formulation is the velocity ratio $C = {V}_\text{peak,t}/{V}_\text{bulk,t}$, which relates the peak throat velocity to the corresponding bulk throat velocity. This ratio allows the model to use a peak stenotic velocity as input when only peak velocity information is available. In a clinical setting, such a peak velocity may be obtained from Doppler ultrasound. In our experiments, $C$ varied between 1.37 and 1.47 across a wide range of measured flow rates. This variation reflects the flattening of the velocity profile for increasing $Re$. However, for a diagnostic tool to be clinically viable, it cannot require a complex, real-time measurement of the full velocity profile to determine the exact $C$ for every stenosis geometry and flow rate. A recent study derived a similar flow-based expression for the SB equation by assuming a fixed velocity ratio $C =$ 1.5 in cerebral venous flow and showed good agreement with CFD simulations \cite{Sidora:2025}. This supports the practical idea of translating between flow rate and peak velocity via a prescribed $C$. Here we tested the validity of simplifying $C$ to a single constant $C_\text{mean}$. Our results indicate that this simplification is remarkably robust. As shown by the narrow gray error bands in Fig.~\ref{fig:p}b, varying $C$ across its entire observed range ($1.37-1.47\ $) resulted in only minor deviations in the predicted pressure drop. While the SB equation effectively ignores the shape of the velocity profile, the MB approach, even with a fixed $C_\text{mean}$, maintains its accuracy across all flow conditions. 

% importance of  correct pressure drop estimation and its relation to our data especially the shaded region (UIV measurements)
Clinical guidelines grade native aortic stenosis (AS) by the mean transvalvular pressure drop: mild disease is characterized with mean gradients roughly below $20 - 25$ mmHg, moderate $25 - 40$ mmHg, and severe stenosis exhibits a gradient above 40 mmHg, where the exact boundaries depend also on the jet velocity and valve area \cite{Baumgartner:2017}. Since pressure drop thresholds directly affect follow-up decisions and the timing of valve intervention, errors in pressure drop estimation can influence disease classification. Such errors may arise from the velocity measurement itself, for example Doppler ultrasound, UIV, or MRI, or from the pressure drop model used to determine the pressure from the velocity. Therefore, improving the physical basis of Bernoulli pressure drop models remains important, particularly in stenotic flows where viscous losses, turbulence production, and pressure recovery may affect the relation between velocity and pressure drop.

The tested $Re$ ($1000-4000\ $) in our experiments lie within the physiological band: recent in-vivo data from healthy subjects report mean $Re$ numbers of about 1000 in the ascending aorta, with peak-systolic values between roughly 4000 and 9000 over the cardiac cycle; in the abdominal aorta and carotids, typical mean values are lower but can still reach several thousand during peak systole \cite{Menon:2024}. In stenotic disease, $Re$ can increase substantially with stenosis severity due to higher velocities in the stenosis area. In addition, as a result of activity or pharmacologic stress, the flow rate and thus $Re$ in the stenosis will increase. As a consequence, the post-stenotic jet becomes more unstable, resulting in additional irreversible losses and thus a higher net pressure drop. In our idealized stenosis model, the clinically relevant $\Delta p = 10 - 15 \ mmHg$ band in Fig.~\ref{fig:p}b corresponds to pipe $Re \approx 2800 - 3900$. This range lies within the peak-systolic $Re$ reported in vivo, highlighting that accurate pressure drop prediction in this band is important for assessing stenosis severity.

% MRI measurements
The MB model was calibrated from direct in-vitro pressure data as a Re-dependent loss coefficient and is formulated to be a function of either flow rate (Eq. \ref{eq:Kexpq}) or peak throat velocity (Eq. \ref{eq:Kmb}). This structure makes the approach naturally compatible with PC-MRI, where the flow rate can be obtained by integrating the velocity field over a segmented lumen. However, the experiments also show that the benefit of a physics-based pressure model can be negated if the stenotic throat is not sufficiently resolved. Across the studied range, the throat bulk velocity is systematically underestimated as the in-plane pixel size increases. This is consistent with PVE due to the velocity near the wall and the relatively small number of pixels that span the cross-section. Because the pressure estimate from the MB model is directly affected by the measured mean velocities, this velocity bias propagates into a strong underestimation of the trans-stenotic pressure drop for large pixel sizes. In other words, when the in-plane pixel size is small (e.g. $0.13 - 0.25 $ mm/px in our study), the MB model in combination with PC-MRI flow rates estimates the pressure drops within approximately $6-13\%$ with respect to reference pressure drop measurements. When the pixel size is increased ($\geq$ 0.50 mm/px), the velocity underestimation dominates and the MB model underpredicts the pressure drop by $20-65\% $. To keep both quantities within a 10$\%$ error range, at least $15-20\ $ pixels across the stenosis lumen are needed for the idealized stenotic model that we used for the current study. This corresponds to a pixel size in the order of one-tenth of the throat radius.

On the other hand, MB-predicted pressure drop using the peak velocity analysis adds an important point to this pixel size effect. While the mean throat velocity shows a strong negative bias with increasing pixel size, the peak throat velocity is robust, with deviations of only a few percent even at the coarsest sampling. As a result, when the MB pressure drop is computed from the peak velocity, the pressure errors are substantially lower and increase more gradually with pixel size (remaining within roughly -3 to -16$\%$ in our experiments). This contrast highlights that the large underestimation observed for the flow-rate-based formulation is primarily driven by the sensitivity of $V_\text{bulk}$ to PVE and segmentation at the wall. In contrast, the peak-based formulation is less affected because $V_\text{peak}$ is governed by the core of the lumen, and is less contaminated by PVE and averaging.

These findings provide an actionable guideline for protocol design: reliable MRI-based pressure estimation requires sufficient sampling across the stenosis lumen to limit partial-volume bias, particularly when the mean velocity used in the MB formulation. While the exact threshold will depend on the signal-to-noise ratio, segmentation strategy, slice thickness, and stenosis geometry, the combined heatmaps provide a concrete pixel-size--error map for both bulk- and peak-based MB implementations in this idealized stenosis model. This map can be used to balance scan time with the accuracy required to obtain clinically meaningful and accurate pressure drops.

%Limitations
%1)
The present study used an idealized circularly symmetric stenosis with one inlet and one outlet under steady flow and rigid-wall conditions. This geometry was chosen as a controlled first step to evaluate the $Re$ dependence of the MB pressure-loss coefficient without the additional complexity of bifurcation flow, vessel compliance, pulsatility, or irregular plaque morphology. The same one-inlet and one-outlet formulation is relevant to several cardiovascular applications, including aortic valve stenosis, coarctation of the aorta, and stenosis located within an individual arterial branch \cite{Baumgartner:2017}. However, the present phantom does not reproduce the full complexity of patient-specific stenoses, and the calibrated coefficients reported here should therefore be regarded as baseline coefficients for this specific rigid geometry rather than universal coefficients for clinical use.

The effect of stenosis geometry on pressure-loss has been investigated in in-vitro studies by Seeley and Young \cite{Seeley:1976}. They showed that pressure losses across arterial stenosis models depend on geometric characteristics such as area reduction, lesion length, eccentricity, and the presence of multiple stenoses. Their results support the use of a pressure-loss relation containing viscous and inertial terms, but also show that the associated coefficients are geometry-dependent. Additionally, they showed that, for severe stenoses, eccentricity had relatively little effect on the overall pressure drop, whereas non-symmetric and irregular stenoses required geometry-specific interpretation of the coefficients \cite{Seeley:1976}. These findings suggest that eccentricity and irregular morphology primarily affect the effective coefficient values, rather than invalidating the use of a pressure-loss relation of the same general form. In the present MB framework, such effects would be reflected through morphology-dependent changes in $k_\text{v}$, $k_\text{t}$, and $C_\text{mean}$.

A second limitation is that the experiments were performed under steady flow conditions, whereas physiological stenotic flows are pulsatile. Steady flow was chosen intentionally to isolate the $Re$-dependent pressure-loss behavior without the additional effects of temporal acceleration, pulsatile inlet flow waveform, and cycle-to-cycle variability. Young and Tsai \cite{Young:1973b} extended their steady stenosis pressure drop model to harmonically oscillating flow by adding an unsteady inertial term. In that formulation, the instantaneous pressure drop is expressed as the sum of a viscous loss term, a turbulence- and separation-related loss term, and an inertial term proportional to the temporal acceleration of the flow. Young and Tsai showed that, for severe stenoses, the viscous and turbulence-related terms can dominate the pressure drop, so that the peak pressure drop may occur close to the peak flow rate and may be approximated reasonably well by a quasi-steady formulation with an additional inertial correction. However, this behavior depends on stenosis severity, geometry, the pulsatile inlet flow waveform, and the resulting flow regime. Therefore, extension of the present MB framework to pulsatile flow would require evaluation of the effective coefficients under pulsatile conditions and, for instantaneous pressure drop prediction, inclusion of the appropriate acceleration-related contribution.

A third limitation is the rigid nature of the FDA nozzle and piping, whereas real arteries are compliant. Chernyavsky et al. \cite{Chernyavsky:2025} investigated pulsatile flow through stenoses embedded in elastic arterial models. Their results showed that prediction accuracy improves when separate coefficient values are used for groups of cases with similar properties, for example, the same vessel elasticity and stenosis severity, rather than using one global coefficient set for all cases. These findings support the interpretation that compliance and pulsatility primarily modify the effective values of $k_\text{v}$ and $k_\text{t}$, and the inertial contribution in unsteady flow, rather than invalidating the reduced-order pressure-loss structure. In the present study, the calibrated coefficients therefore represent a rigid-wall, steady flow limit.

A future extension of this approach could therefore be the development of a coefficient library for $k_\text{v}$, $k_\text{t}$, and $C_\text{mean}$ that accounts for both stenosis morphology and relevant physiological conditions. Direct pressure-based recalibration of these coefficients for each individual patient would not be practical in routine clinical use. Instead, such a library could be generated from controlled experiments or CFD simulations covering a range of stenosis geometries, flow waveforms, and vessel compliance values. For new cases, coefficients could then be selected or interpolated from the established library, based on clinical and imaging input. This would preserve the reduced-order nature of the MB formulation while accounting for morphology- and condition-dependent pressure-loss behavior. Recent work by Hilhorst et al. \cite{Hilhorst:2026} supports this direction, showing that geometry-dependent pressure-loss behavior can be learned from CFD-derived data across a range of coronary stenosis morphologies using statistical shape descriptors. This extension has not been demonstrated in the present study and remains a necessary step before clinical translation.

Finally, the present study does not include clinical measurements. The experiments were performed in a controlled rigid phantom with a single inlet and outlet, which allowed the pressure-loss behavior to be studied under well-defined conditions, but did not reproduce the full complexity of physiological stenotic flow. Validation using independent clinical or patient-specific datasets, ideally with direct pressure measurements where available, will be necessary before the method can be used for clinical decision-making. The coefficient-library concept described above has not been demonstrated in the present study and remains a necessary step before clinical translation.
% MB formulation and conclusion

To conclude, this study provides a controlled baseline evaluation of a $Re$-dependent modified Bernoulli model as a pressure-loss formulation for stenotic flows. This model was calibrated using direct pressure drop measurements and velocity data obtained from ultrasound imaging velocimetry and PC-MRI. Within the investigated steady flow, one-inlet and one-outlet phantom configuration, the MB formulation reduced the discrepancy with the measured pressure drops compared with the SB and EB equations. This improvement is attributed to the inclusion of $Re$-dependent loss behavior, which allows the model to represent the changing balance between viscous losses and geometry-dependent inertial, separation, and turbulence-related losses. The MRI analysis further showed that pressure estimates are sensitive to spatial sampling and effective image resolution. Increasing the nominal in-plane pixel size led to systematic underestimation of throat velocity and, consequently, pressure drop. For the smallest investigated in-plane pixel size, the MB pressure estimates remained closer to the experimental pressure measurements. The peak throat velocity was less sensitive to the in-plane pixel size than the bulk throat velocity. It suggests that a peak-velocity-based form of the MB model, combined with an appropriate estimate of $C_\text{mean}$, may be useful when only peak stenotic velocity is available for estimating the trans-stenotic pressure drop.

\section*{Data availability}

The datasets supporting the conclusions of this article are included within the article and its Supplementary Information files. The Supplementary Information contains additional theoretical details and tabular data supporting the Methods and Results. Dataset 1 contains the experimental results for pressure drop and velocity measurements. Dataset 2 contains the FDA nozzle geometry as an STL file.

\bibliography{sample}

\section*{Funding }

This project was supported by TU Delft Faculty of Mechanical Engineering through a ME Cohesion grant awarded to S.P. and W.H. We acknowledge Prof.\ Christian Poelma for valuable discussions on the UIV data analysis and for providing access to the UIV measurement facilities.

\section*{Author contributions statement}

\noindent
A.A.: Conceptualization, Data curation, Formal analysis, Investigation, Methodology, Project administration, Resources, Software, Validation, Visualization, Writing – original draft, Writing – review $\&$ editing.
J.T.P.: Conceptualization, Methodology, Project administration, Supervision, Writing – review $\&$ editing.
S.P.: Conceptualization, Funding acquisition, Methodology, Project administration, Resources, Supervision, Visualization, Writing – review $\&$ editing.
W.H.: Conceptualization, Funding acquisition, Methodology, Project administration, Resources, Software, Supervision, Validation, Visualization, Writing – review $\&$ editing.

\section*{Competing interests}
The authors declare no competing interests.

\section*{Additional information}
\textbf {Supplementary Information} The online version contains supplementary material.

\noindent
\textbf {Correspondence} and requests for materials should be addressed to S.P. and W.H.
 
\end{document}